\documentclass[fleqn,usenatbib, nofootnoteinbib]{mnras}
\usepackage[utf8]{inputenc}  
\usepackage[T1]{fontenc}
\usepackage{graphicx,psfrag}
\usepackage{mathrsfs}
\usepackage{amsmath,amsfonts,amssymb,pifont,gensymb}
\usepackage{multirow,enumerate}
\usepackage{comment,hyperref}
\usepackage{color}
\usepackage{acronym}
\usepackage{xspace}
\usepackage[normalem]{ulem}
\usepackage{mathtools}
\usepackage{subfigure}
\usepackage{makecell}
\usepackage{newtxtext,newtxmath}
\usepackage{changepage}
\usepackage{appendix}
\usepackage{lipsum}

\newcommand{\golum}{\textsc{GOLUM}\:}
\newcommand{\boldTheta}{\mathbf{\Theta}}
\newcommand{\boldLambda}{\mathbf{\Lambda}}
\newcommand{\boldPhi}{\mathbf{\Phi}}
\newcommand{\Le}{\mathrm{L}}
\newcommand{\U}{\mathrm{U}}
\newcommand{\lenshypo}{\mathcal{H}_{\mathrm{L}}}
\newcommand{\unlhypo}{\mathcal{H}_{\mathrm{U}}}
\newcommand{\clu}{\mathcal{C}^{\mathrm{L}}_{\mathrm{U}}}
\newcommand{\dd}{\mathrm{d}}
\newcommand{\bvarTheta}{\boldsymbol{\vartheta}}
\newcommand{\ifo}{\mathrm{ifo}}
\newcommand{\blangle}{\big\langle}
\newcommand{\brangle}{\big\rangle}
\newcommand{\neff}{N_{\mathrm{eff}}}

\usepackage{geometry}
\title[Improving Distributed Joint Parameter Estimation for Strongly-Lensed GWs]{The Return of GOLUM: Improving Distributed Joint Parameter Estimation for Strongly-Lensed Gravitational Waves}
\author[J. Janquart et al.]{
Justin Janquart$^{1, 2}$\thanks{E-mail: j.janquart@uu.nl},
Haris K.$^{1, 2}$,
Otto A. Hannuksela$^{3}$,
and  Chris Van Den Broeck$^{1, 2}$
\\
$^{1}$Nikhef – National Institute for Subatomic Physics, Science Park, 1098 XG Amsterdam, The Netherlands\\
$^{2}$Institute for Gravitational and Subatomic Physics (GRASP), Department of Physics, Utrecht University, Princetonplein 1, 3584 CC Utrecht, The Netherlands \\
$^{3}$ Department of Physics, The Chinese University of Hong Kong, Shatin, N.T., Hong Kong
}

\date{\today}

\pubyear{2021}

\begin{document}

\maketitle

\begin{abstract}
\noindent
Owing to the forecasted improved sensitivity of ground-based gravitational-wave detectors, new research avenues will become accessible. This is the case for gravitational-wave strong lensing, predicted with a non-negligible observation rate in the coming years. However, because one needs to investigate all the event pairs in the data, searches for strongly-lensed gravitational waves are often computationally heavy, and one faces high false-alarm rates. In this paper, we present upgrades made to the \textsc{GOLUM} software, making it more reliable while increasing its speed by re-casting the look-up table, imposing a sample control, and implementing symmetric runs on the two lensed images. We show how the recovered posteriors have improved coverage of the parameter space and how we increase the pipeline's stability. Finally, we show the results obtained by performing a joint analysis of all the events reported until the GWTC-3 catalog, finding similar conclusions to the ones presented in the literature.
\end{abstract}

\section{Introduction}
Since the first detection of a binary black hole (BBH) merger through gravitational wave (GW) observation~\citep{TheLIGOScientific:2016wfe}, the detectors have been upgraded continually, leading to the detection of close to 100 compact binary coalescences (CBCs) reported after the third observing run~\citep{LIGOScientific:2021djp}. These detections enabled us to study the population of compact binary mergers~\citep{2021arXiv211103634T}, cosmology~\citep{2021arXiv211103604T}, and test general relativity~\citep{2021arXiv211206861T}. In the coming years, the LIGO~\citep{TheLIGOScientific:2014jea} and Virgo~\citep{TheVirgo:2014hva} detectors should be upgraded further, improving their sensitivity even more. In addition, the KAGRA~\citep{Somiya:2011np, Aso:2013eba, Akutsu:2018axf, Akutsu:2020his} detector should join the detector network soon, and LIGO-India is under construction~\citep{LigoIndia}. The combination of increased detector number and improved sensitivity will open the way to new scientific avenues. 

A new type of observation, forecast for the coming years~\citep{Ng:2017yiu, Oguri:2018muv, Wierda:2021upe}, is that of strongly-lensed GWs. Strong lensing happens when the characteristic wavelength is larger than the typical size of the lens (often described by its Schwarzchild radius), which leads to multiple images with the same frequency evolution. The different images are magnified, phase shifted, and delayed in time~\citep{Ohanian1974, Degushi1986, Wang:1996as, Nakamura1998, Takahashi:2003ix, Dai:2017huk, Ezquiaga:2020gdt, Oguri:2018muv, Liu:2020par, Lo:2021nae, Janquart:2021qov}. If the lens is a galaxy, the images are typically separated from minutes to months~\citep{Dai:2016igl, Ng:2017yiu, Li:2018prc, Oguri:2018muv, Wierda:2021upe}. If it is a galaxy cluster the time delay can grow to years~\citep{Smith:2017mqu, Smith:2018gle, Smith:2019dis, Robertson:2020mfh, Ryczanowski:2020mlt}. If not accounted for, lensing effects can lead to biased results. The magnification of the GW leads to a modified amplitude which can bias the luminosity distance~\citep{Dai:2016igl, Ng:2017yiu, Pang:2020qow}. When higher-order modes (HOMs) are present, the overall phase-shift can lead to biased posteriors~\citep{Dai:2017huk, Ezquiaga:2020gdt,2022arXiv220206334V, Janquart:2021nus}. Consequently, if a lensed event has a significant HOM contribution, strong lensing can be identified more easily~\citep{Wang:2021kzt, Lo:2021nae, 2022arXiv220206334V, Janquart:2021nus}. 
Other types of lensing are also possible. If the typical size of the lens is comparable to the GW wavelength, one gets frequency-dependent beating patterns, often referred to as microlensing~\citep{Takahashi:2003ix, Cao:2014oaa, Lai:2018rto, Christian:2018vsi, Singh:2018csp, Hannuksela:2019kle, Meena:2019ate, Pagano:2020rwj, Cheung:2020okf, Kim:2020xkm, Wright:2021cbn}.

Strongly-lensed events would lead to new possibilities to probe fundamental physics. Because one gets multiple similar images, the effect is similar to an enhanced detector network made by the sum of the detectors observing each image. An extended network greatly improves the sky localization for BBHs~\citep{Hannuksela:2020xor, Lo:2021nae, Janquart:2021qov}, would allow for better probes of GW polarization~\cite{Goyal:2020bkm, MaganaHernandez:2022ayv}, and to better probe higher-order modes should they be present~\cite{Janquart:2021nus}. In addition, it would be possible to match the lensed GW data with electromagnetic (EM) data~\citep{Hannuksela:2020xor, Wempe:2022iop}, leading to sub-arcsecond precision localization capabilities~\citep{Hannuksela:2020xor, Wempe:2022iop}. For galaxy lenses, finding the corresponding lens-source system in the EM band generally requires the detection of four images. For galaxy-cluster lenses, thanks to their scarcity,  the same could be obtained using only two images~\citep{Smith:2018gle, Ryczanowski:2020mlt, Sereno:2011ty, Yu:2020agu}. Finding an EM counterpart to a GW-lensed event would also confirm the lensed nature of the event. Observing the two channels is also of interest for other tests of fundamental physics~\citep{Baker:2016reh, Fan:2016swi}. The milli-second precision offered by strongly lensed events is complementary with the information given by the EM data, leading to the possibility of performing precision cosmography if a joint detection was made~\citep{Sereno:2011ty, Liao:2017ioi, Cao:2019kgn, Li:2019rns}. Even on its own, GW strong lensing offers interesting perspectives. 

GW lensing searches are already ongoing~\citep{Hannuksela:2019kle, LIGOScientific:2021izm, LIGOScientific:2023bwz}. So far, no confident detection has been reported, even if some intriguing candidates have been identified~\citep{Dai:2020tpj}. The main philosophy behind strong-lensing searches is identifying event pairs with the same intrinsic parameters, and originating from the same sky location. Several approaches have been developed to address various challenges. An intuitive approach, called posterior overlap, compares the posteriors obtained from unlensed searches~\citep{Haris:2018vmn}. One takes the posteriors for (a subgroup of) the binaries' parameters, makes a KDE reconstruction, and computes how (di)similar they are. For a lensed pair, one expects the two to overlap significantly. This method is relatively fast. However, overlaps can happen by chance, leading to risks of false-alarms~\citep{Wierda:2021upe, Caliskan_2022, Janquart_2022}. A significantly more accurate method is joint parameter estimation, where two data streams are analyzed simultaneously, assuming they have the same intrinsic parameters and are linked through their relative magnification, time delay, and overall phase shift~\citep{Liu:2020par, Lo:2021nae}. A faster alternative was presented in~\citet{Janquart:2021qov}. The idea is to distribute the runs rather than doing them simultaneously, hence analyzing the first image under the lensed hypothesis and using the results to analyze the second image. This leads to a faster parameter estimation scheme, equivalent under the lensed hypothesis. This approach is implemented in a framework called \textsc{GOLUM}~\citep{Janquart:2021qov, Golum_git}. In this work, we show developments made to this framework to enhance its speed and reliability.

One of the main issues faced when searching for strongly-lensed  GWs is the rapidly growing false alarm probability (FAP)~\citep{Wierda:2021upe, Caliskan_2022, Janquart_2022} related to the fast increase in the number of pairs to analyze~\citep{Ng:2017yiu}. Indeed, when searching for strong lensing, one should consider all the pairs of events one can make from the single events, giving a quadratically growing number of pairs. In past analyses~\citep{LIGOScientific:2021izm, LIGOScientific:2023bwz}, the technique has been to analyze the data in multiple steps, going from the fastest and least accurate method (posterior overlap) to the most accurate and computationally-heavy one. However, in upcoming runs, the false alarms given by posterior overlap could lead to many events to follow up using joint parameter estimation. It could become impossible to follow, requiring an enhanced intermediate step. Including a lens model in the detection statistic is one way to decrease the false-alarm risk~\citep{Haris:2018vmn, Wierda:2021upe, Janquart_2022}. However, using the wrong model can mean one misses a genuinely lensed event~\citep{Janquart_2022}. The first way to decrease the FAP is to use more accurate methods~\citep{Janquart_2022}. Unfortunately, such methods are generally computationally heavier than less accurate ones, leading to issues when trying to follow up. It has been suggested in~\citet{Janquart_2022} that a method like \golum could bypass the most expensive step  -- the first image run -- under the condition that HOMs are not too strong. Then, the analysis of the first image could be skipped, opening the door to the direct use of a more precise methodology.

In this work, we demonstrate some improvements added to the \golum software to address some of the challenges still present in the strong lensing data analysis. The work is structured as follows. In Sec.~\ref{sec:theory_strong_lensing}, we present the basics of strong lensing. In Secs.~\ref{sec:bayes_analysis_GW} and~\ref{sec:golum_overview}, we present how one can perform Bayesian inference for strongly-lensed GWs and how it can be tweaked to lead to a faster method. Secs.~\ref{sec:gol_speed}, \ref{sec:samp_cont}, \ref{sec:low_lat_gol} and \ref{sec:symm_gol} outline the different updates implemented in \golum. We then compare the updated pipeline with joint parameter estimation in Sec.~\ref{sec:compa_jpe}, and show our results for the GW data released by the LVK in Sec.~\ref{sec:o3_data}. Finally, we give our conclusions in Sec.~\ref{sec:conclusions}.

\section{Strongly-lensed Gravitational Waves}
\label{sec:theory_strong_lensing}

Strong lensing leads to several possible detectable images. These images are modified in three ways. First, they can be (de)magnified, translated by a magnification factor $\mu_j$. Formally, the magnification is the inverse of the determinant of the lensing Jacobian matrix~\citep{Schneider:1992, Haris:2018vmn}. Additionally, the image can undergo an overall phase shift~\citep{Dai:2017huk, Ezquiaga:2020gdt}, represented by a Morse factor $n_j$. It can only take three different values, making for three types of lensed images. One has type I images when $n_j = 0$. This happens when the image passes through the minimum of the Fermat potential and is equivalent to no phase shift. If the image passes through a saddle point, $n_j = 0.5$, and the image is of Type II. Finally, if $n_j = 1$, one has a type III image, and the lensed GW passes through a maximum of the Fermat potential. When HOMs are present, the overall phase-shift induced by the Morse factor for a type II image cannot be mistaken for change in the binary's phase value. Therefore, it can lead to the image identification and smoking-gun evidence for strong-lensing~\citep{Ezquiaga:2020gdt, Janquart:2021nus, 2022arXiv220206334V}. Since different images follow a different geometric path, they have an additional time delay $t_j$. It depends on the nature of the lens. For a galaxy lens, one expects minutes to month time delays~\citep{Takahashi:2003ix, Dai:2017huk, Ng:2017yiu, Oguri:2018muv}, while it can reach years for galaxy cluster lenses~\citep{Smith:2017mqu, Smith:2018gle, Smith:2019dis, Robertson:2020mfh, Ryczanowski:2020mlt}.

Accounting for these different effects, the lensed $h_{\mathrm{L}}^j$ and unlensed $h_{  \mathrm{U}}$ waveforms for an image $j$ are linked as
\begin{equation}
    \tilde{h}^{j}_{\mathrm{L}}(f; \boldTheta, \boldLambda_j) = \sqrt{\mu_j}\tilde{h}_{\mathrm{U}}(f; \boldTheta) e^{\big(2 i\pi f t_j - i\pi n_j \mathrm{sign}(f)\big)} \, ,
\end{equation}
where $\boldTheta$ represents the usual BBH parameters, $\boldLambda_j = \{\mu_j, t_j, n_j \}$ represents the lensing parameters, and the tilde specifies we are working in the frequency domain. 

Considering solely geometric optics\footnote{We do not consider inference from micro-lensing or effects such as the lifting of the mass-sheet degeneracy~\citep{Cremonese:2021puh}.}, the magnification and the time delay are generally not measurable on their own. For one image, the magnification can be absorbed in an observed luminosity distance 
\begin{equation}\label{eq:obs_dl}
    D_{L}^{\mathrm{obs}, j} = \frac{D_L}{\sqrt{\mu_j}}\, ,
\end{equation}
where $D_L$ is the source luminosity distance. The time delay can also be absorbed in an observed time of coalescence 
\begin{equation}\label{eq:obs_tc}
  t_c^{\mathrm{obs}, j} = t_c + t_j \, ,
\end{equation}
where $t_c$ is the unlensed time of coalescence. Formally, the Morse factor cannot be absorbed in an effective phase as the presence of HOMs would lift such a degeneracy~\citep{Ezquiaga:2020gdt, 2022arXiv220206334V, Janquart:2021nus}. Due to these degeneracies, one studies the images by pairs and relates them via the relative lensing parameters $\boldPhi = \{\mu_{21}, t_{21}, n_{21}\}$, with
\begin{align}\label{eq:rel_params}
    t_c^{\mathrm{obs}, 2} &= t_c^{\mathrm{obs}, 1} + t_{21} \, , \nonumber \\
    D_L^{\mathrm{obs}, 2} &= \sqrt{\mu_{21}} D_L^{\mathrm{obs}, 1} \, , \\
    n_2 &= n_1 + n_{12}  \, . \nonumber
\end{align}
With these parameters, it is easy to express the waveform for one image as a function of the waveform of another one. 

An alternative possibility for the analysis is to sample directly the observed parameters (apparent luminosity distance, Morse factor, and coalescence time) for the two images. This does not account entirely for the lensing hypothesis since it does not link the observed parameters. However, this can be easier when including population models, and the correlation is added in the post-processing step when the models are added~\citep{Lo:2021nae}. This parametrization has been added to the \golum framework to incorporate selection effects in future studies.

\section{Bayesian Analysis for Strongly-Lensed Gravitational Waves}
\label{sec:bayes_analysis_GW}
The search for strong lensing is a Bayesian model selection problem. One has to decide whether it is more likely to observe the data under the lensed or the unlensed hypothesis. In GW lensing, one analyzes two data streams jointly to determine whether the GWs present are lensed images of each other. One of the data streams can be written
\begin{equation}\label{eq:data_stream}
    d_j = n_j(t) + h_{\mathrm{H}}^{j}(\mathbf{\Xi}_j) \, ,
\end{equation}
where $n_j(t)$ is the noise realization for the stretch of data, and $h_{\mathrm{H}}^{j}(\mathbf{\Xi_j})$ is the GW signal buried in the noise. It can either be a lensed image ($h_{\mathrm{H}}^{j} = h_{\Le}^{j}$ and $\mathbf{\Xi}_j = \{\boldTheta, \boldLambda_j\}$) of a multiplet or an unlensed GW event ($h_{\mathrm{H}}^{j} = h_{\U}^j$ and $\mathbf{\Xi}_j = \boldTheta_j$).  

The goal is then to determine if it is more likely to be in the lensed hypothesis $\lenshypo$, hence the two observed GWs are images of each other described by $\{\boldTheta, \boldLambda_1, \boldLambda_2\}$, or to be in the unlensed hypothesis $\unlhypo$, thus the two GWs are independent and described by $\{\boldTheta_1, \boldTheta_2\}$. Typically, the comparison between the two hypotheses is done using the Odds ratio
\begin{equation}\label{eq:odds_ratio}
    \mathcal{O}^{\lenshypo}_{\unlhypo} = \frac{P(\lenshypo | d_1, d_2)}{P(\unlhypo | d_1, d_2)} = \frac{P(\lenshypo)}{P(\unlhypo)}\frac{P(d_1, d_2 | \lenshypo)}{P(d_1, d_2 | \unlhypo)} \,,
\end{equation}
where we have just developed the Odds ratio's definition using Bayes theorem. $P(\lenshypo)/P(\unlhypo)$ is the prior odds and tells how likely it is to be in one of the two hypotheses before collecting any data. While it may not always be the correct thing to do~(Hannuksela et al, in prep.), it often is disregarded, and only the second ratio, usually called the Bayes factor, is used to decide whether an event is lensed or not. In this work, we will use the conventions set by~\citet{Lo:2021nae, Janquart:2021qov} and call the ratio of evidence
\begin{equation}\label{eq:coherence_ratio}
    \clu = \frac{P(d_1, d_2 | \lenshypo)}{P(d_1, d_2 | \unlhypo)} \, ,
\end{equation}
the coherence ratio when it does not account for selection effects\footnote{See~\citet{Lo:2021nae} for a detailed discussion on the subject. In this work, the different quantities neglect them because they can simply be added in a post-processing step.}. The term Bayes factor refers to the case where selection effects are included. 

Under the lensed hypothesis and neglecting selection effects, the joint evidence for two data streams 
\begin{align}\label{eq:joint_evidence_lensed}
p(d_1, d_2 | \lenshypo) = \int &p(d_1 | \boldTheta, \boldLambda_1) p(d_2 | \boldTheta, \boldLambda_2) \nonumber \\
&\times p(\boldTheta, \boldLambda_1, \boldLambda_2) \dd\boldTheta\dd\boldLambda_1 \dd\boldLambda_2 \, ,
\end{align}
where $p(\boldTheta, \boldLambda_1, \boldLambda_2)$ is the joint prior on the binary parameters and the lensing parameters, and $p(d_j | \boldTheta, \boldLambda_j)$ with $j = \{1, 2\}$ are the individual likelihoods. 

Eq.~\eqref{eq:joint_evidence_lensed} can be recast by using the observed parameters~\eqref{eq:obs_dl},~\eqref{eq:obs_tc} and the relative lensing parameters~\eqref{eq:rel_params} 
\begin{align}\label{eq:joint_ev_rel_par}
    p(d_1, d_2 | \lenshypo) = \int &p(d_1 | \bvarTheta) p(d_2 | \bvarTheta, \boldPhi) \nonumber \\
&\times p(\bvarTheta, \boldPhi) \dd\bvarTheta\dd\boldPhi \, ,
\end{align}
where $\bvarTheta$ represents the binary parameters with the Morse factor and the observed luminosity distance and time delay for the first image. $\boldPhi$ is the set of relative lensing parameters linking the two images. Eqs.~\eqref{eq:joint_evidence_lensed} and~\eqref{eq:joint_ev_rel_par} are equivalent and represent simply a different way of expressing the waveforms and the link between them. This form has the main advantage of being easier to distribute in the \golum framework (see Sec.~\ref{sec:golum_overview}). 

Under the unlensed hypothesis, the two data streams are simply uncorrelated. Therefore,
\begin{align}\label{eq:unl_evidence}
    p(d_1, d_2 | \unlhypo) &= \int p(d_1 | \boldTheta_1) p(d_2 | \boldTheta_2) p(\boldTheta_1, \boldTheta_2) \dd\boldTheta_1 \dd\boldTheta_2 \nonumber \\
    &= \int p(d_1 | \boldTheta_1)p(\boldTheta_1)\dd\boldTheta_1 \int p(d_2 | \boldTheta_2)p(\boldTheta_2)\dd\boldTheta_2 \nonumber \\
    &= p(d_1 | \unlhypo) p(d_2 | \unlhypo) \, .
\end{align}
Here, the priors are unrelated, and we directly get the product of the evidence under the unlensed hypothesis. This can, in principle, directly be obtained using the unlensed data analysis, generally performed by the LVK~\citep{LIGOScientific:2021djp}. 

Using Eqs.~\eqref{eq:joint_evidence_lensed} and~\eqref{eq:unl_evidence} (or alternatively Eqs.~\eqref{eq:joint_ev_rel_par} and~\eqref{eq:unl_evidence}) one can express the coherence ratio~\eqref{eq:coherence_ratio}. Evaluating the coherence ratio in this form is the task performed by joint parameter estimation tools~\citep{Liu:2020par, Lo:2021nae}, where one formally evaluates the joint likelihood. To make the comparison with our faster approach possible, we also added a joint likelihood inference tool in the \golum package~\citep{Golum_git}. 

\section{GOLUM's Trick}
\label{sec:golum_overview}

Distributed joint parameter estimation makes use of the possibility to re-express the joint evidence~\eqref{eq:joint_ev_rel_par}\footnote{The same developments work when parametrizing the evidence as in Eq.~\eqref{eq:joint_evidence_lensed} but, for simplicity, we focus on one way to cast the problem. The results for the other parametrization are obtained similarly.} as~\citep{Janquart:2021qov}
\begin{equation}\label{eq:distributed_joint_evidence}
    p(d_1, d_2 | \lenshypo) = p(d_2 | d_1, \lenshypo)p(d_1 | \lenshypo) \, ,
\end{equation}
a product of the evidence for the first image only under the lensed hypothesis, hence accounting for the Morse factor, and the conditional evidence of the second data set given the observation of the first. Mathematically, 
\begin{align}\label{eq:conditional_ev_detail}
    p(d_2 | d_1, \lenshypo) = \int &\bigg[\int p(d_2 | \bvarTheta, \boldPhi, \lenshypo) p(\bvarTheta | d_1, \lenshypo) \dd\bvarTheta\bigg] \nonumber \\ & \times p(\boldPhi | \lenshypo)\dd\boldPhi \, .
\end{align}
Now, $p(\bvarTheta | d_1, \lenshypo)$ is the probability of having a given observed parameter in the first image given its data. It is nothing less than the posteriors one would get by analyzing the first image under the lensed hypothesis. Eq.~\eqref{eq:conditional_ev_detail} has the main advantage of leading to a faster evaluation than the full likelihood, expressing it in terms of a ``marginalized'' likelihood 
\begin{equation}\label{eq:cond_ev_marg_likeli}
    p(d_2 | d_1, \lenshypo) = \int L(\boldPhi | \lenshypo)p(\boldPhi | \lenshypo) \dd\boldPhi \, ,
\end{equation}
where 
\begin{equation}\label{eq:marg_likeli}
    L(\boldPhi | \lenshypo) = \big\langle p(d_2 | \bvarTheta, \boldPhi, \lenshypo) \big\rangle_{p(\bvarTheta | d_1, \lenshypo)} \, .
\end{equation}
While being fully equivalent to the joint likelihood, Eq.~\eqref{eq:cond_ev_marg_likeli} has the advantage of having a faster evaluation. Indeed, since we use the posterior of the first image, the parameters are already sampled around the correct region of the parameter space, making for a better convergence. Moreover, these posteriors are known in advance, making it possible to generate a look-up table before the run, avoiding the computationally-expensive step of waveform generation during the sampling process. 

One issue at this stage is that we have not impacted the second image observation on the posteriors of the first image. This can be done by applying an additional reweighting step using~\citep{Janquart:2021qov}
\begin{equation}\label{eq:simple_reweighitng}
    p(\bvarTheta, \boldPhi | d_1, d_2) \propto \frac{p(d_2 | \bvarTheta, \boldPhi)}{L(\boldPhi)} p(\bvarTheta | d_1)p(\boldPhi | d_1, d_2) \, ,
\end{equation}
where only $p(d_2 | \bvarTheta, \boldPhi)$ is an unknown since the other terms are the posteriors for the first and second image runs, and the denominator is the lensed likelihood evaluated for the various samples during the second image run. 

Using this method, Ref.~\citep{Janquart:2021qov} showed the possibility of getting accurate posteriors with a run time of about 1 CPU hour for the second image. In the end, the time needed to analyze the first image -- corresponding to the tome needed to analyze a single unlensed BBH -- is the dominating factor in the analysis. It is a major improvement compared to joint parameter estimation. However, using the \golum approach in various works has shown that it sometimes has shortcomings due to assumptions made in the derivation. In the coming sections, we show how to overcome some of these. So we update \golum to get a pipeline with (at least) the same speed but an improved convergence and reliability. 

\section{Making GOLUM more efficient}
\label{sec:gol_speed}

 The lookup table presented in~\citet{Janquart:2021qov} gives a faster way to evaluate the likelihood for the second image analysis. The main idea is to pre-compute the weighted inner products for each sample coming from the first image run for each detector. Then, during the nested sampling run, it suffices to correct these sums by the relative lensing parameters and put together the contribution of the different detectors. The evidence for the second event can be written as~\citep{Janquart:2021qov}
\begin{align}\label{eq:old_lookup}
    2\ln(p(d_2 | \bvarTheta, \boldPhi)) = \sum_{\ifo} \Bigg[ &\blangle d_2^{\ifo} | d_2^{\ifo} \brangle + \frac{1}{\mu_{21}} \blangle h_2^{\ifo}(\bvarTheta) | h_2^{\ifo}(\bvarTheta) \brangle \nonumber \\
    & -\frac{2 e^{i\pi n_{21}}}{\sqrt{\mu_{21}}} \blangle d_2^{\ifo} | e^{2 i \pi f t_{21}}h_2^{\ifo}(\bvarTheta) \brangle\Bigg] \, ,
\end{align}
where ``$\ifo$'' runs over the detectors in the network, and the weighted inner product
\begin{equation}\label{eq:weighted_inner}
    \blangle k | l \brangle = 4\int_{-\infty}^{+\infty} \frac{\tilde{k}(f) \tilde{l}^{*}(f)}{S_n(f)} \, .
\end{equation}
${}^*$ means the complex conjugate and $S_n$ is the power spectral density (PSD). This is an intuitive way of constructing the lookup table as it is a direct translation of the maths one would do. In the frequency domain, one needs to make one such table for each possible value of the Morse factor difference. So, we compute a value to store for each time sample in each detector for a given Morse factor difference and repeat for all the possible differences. Two issues were encountered with this version of the lookup table: it is memory extensive and not optimal. To reduce both of these issues, it suffices to realize the lensing parameters (except the time delay) are just prefactor of the various sums
\begin{align}\label{eq:lookup_recast}
    2\ln(p(d_2 | \bvarTheta, \boldPhi)) = & \sum_{\ifo} \blangle d_2^{\ifo} | d_2^{\ifo} \brangle  + \frac{1}{\mu_{21}} \sum_{\ifo} \blangle h_2^{\ifo}(\vartheta) | h_2^{\ifo}(\vartheta) \brangle \nonumber \\ 
    &-\frac{2 e^{i\pi n_{21}}}{\sqrt{\mu_{21}}} \sum_{\ifo} \blangle d_2^{\ifo} | e^{2 i \pi f t_{21}}h_2^{\ifo}(\vartheta) \brangle \, .
\end{align}
The first term in this equation is independent of the parameters as it is the inner product of the data with itself. This can be computed once and for all at the start and used as a correction factor later on. For the two other terms, instead of storing one value for each detector and each time delay, we can save the sum over the detectors directly, reducing the memory. In addition, since we do not need to perform the various sums for each sample in each iteration during the sampling process, we also reduce the number of operations and, thus, the computation time. Using this version of the lookup table and more efficient coding, we reduced the second image run by a factor of five on average\footnote{The comparison in computational speed is done using the same  \emph{Intel(R) Core(TM) i7-9750H CPU @ 2.60GHz} processor as in~\citet{Janquart:2021qov}.}.

Finally, since $t_{21}$ is convolved with the data, one has to compute each sum for each time step in the data stretch. However, this makes for a large memory consumption, especially for longer-duration, hence lower-mass, events. Computing the sums for so many time samples is not necessary. Indeed, one typically has quite a good prior knowledge of the time delay, and the time-prior is often very restricted (typically a uniform distribution in $t_c^{\mathrm{measured}} \pm 0.2s$ at most) and values outside of the priors are unexplored. Therefore, we updated the lookup table to take in the time-prior and pre-compute the different sums only for times included in the prior. The memory gain depends on the data duration, but it is a minimum factor of 10, making \golum more accessible for more extensive studies, lower-mass systems and lower-memory systems.

The combination of the two improvements also enables one to easily use a higher number of samples from the first image posterior without drastically increasing the computational time and memory requirements. In turn, it can be used to improve the analysis' accuracy.

\section{Sample Control}
\label{sec:samp_cont}
When passing from the first to the second image run, it is common to sub-sample the posteriors to have a faster run. This approach can be applied without hurdle as long as the samples cover the parameter space. Afterward, each sample is used in Eq.~\eqref{eq:marg_likeli}, where it contributes to a weight in the sum. Therefore, each selected sample does not have the same final importance and contributes differently to the analysis. As for a reweighting process, we can compute the number of effective samples~\citep{Elvira_2018, Farr_2019, Golomb:2021tll, Payne:2019wmy} we have when drawing $N$ samples from the first image run.

The generic expression for the number of effective samples $\neff$ for a process with weights $w_j$ is~\citep{Elvira_2018, Payne:2019wmy}
\begin{equation}\label{eq:generic_n_eff}
    \neff = \frac{\bigg(\sum_{j = 1}^{N} w_j \bigg)^2}{\sum_{j = 1}^N w_j^2}\, ,
\end{equation}
where $N$ is the total number of samples selected. \\For \golum,
\begin{equation}\label{eq:n_eff_gol_gen}
    w_{j} = p(d_2 | \bvarTheta_j, \boldPhi) \, .
\end{equation}
In Eq.~\eqref{eq:n_eff_gol_gen}, the weights do not depend only on the parameters we are sub-sampling but also on the parameters we still need to infer. Therefore, we cannot directly evaluate the number of effective samples. Still, one can estimate the number of effective samples for a given number of total samples as
\begin{equation}\label{eq:n_eff_est_gol}
    \neff \simeq \frac{1}{N_{\boldPhi}} \sum_{i = 1}^{N_\boldPhi} \Bigg[ \frac{\big(\sum_{ j = 1}^{N} p(d_2 | \bvarTheta_j, \boldPhi_i)\big)^2}{\sum_{ j = 1}^{N} p(d_2 | \bvarTheta_j, \boldPhi_i)^2}\Bigg] \, ,
\end{equation}
where we pre-select $N_\boldPhi$ lensing parameters from the prior and see the number of effective samples obtained on average for the different sets of prior values. Even if it provides an approximate number of effective samples, this relation is already useful to guide the number of samples one should select from the initial distributions to have a given number of effective samples in the analysis. So, by iteratively increasing $N$ in Eq.~\eqref{eq:n_eff_est_gol}, one can find the number of samples such that the estimated $\neff$ is bigger or equal to a target value. Typically, having about or more than 1000 effective samples is enough for a proper convergence of the likelihood. However, the corresponding number of samples to take from the posteriors of the first image is not necessarily 1000. For nicely lensed images (high SNR, well-behaved noise), taking around 1000-1500 samples is enough. However, if one wants stability when analyzing unlensed events or fainter signals, the number of samples to take from the prior can grow to much more samples. Fig.~\ref{fig:Nsamp_Likeli_stab} shows the stability as a function of the number of selected samples. Using 1000 samples leads to a likelihood close to the converged value. Using 2000 samples already leads to much better and more stable results. The first case typically has a corresponding effective number of samples between 700 and 900, while the second case always has more than 1000 effective samples. So, from these experiences, \golum gives stable evidence evaluation when using more than 1000 effective samples. For a typical lensed event pair, using 2000 samples ensures this condition.

\begin{figure}
    \centering
    \includegraphics[keepaspectratio, width=0.5\textwidth]{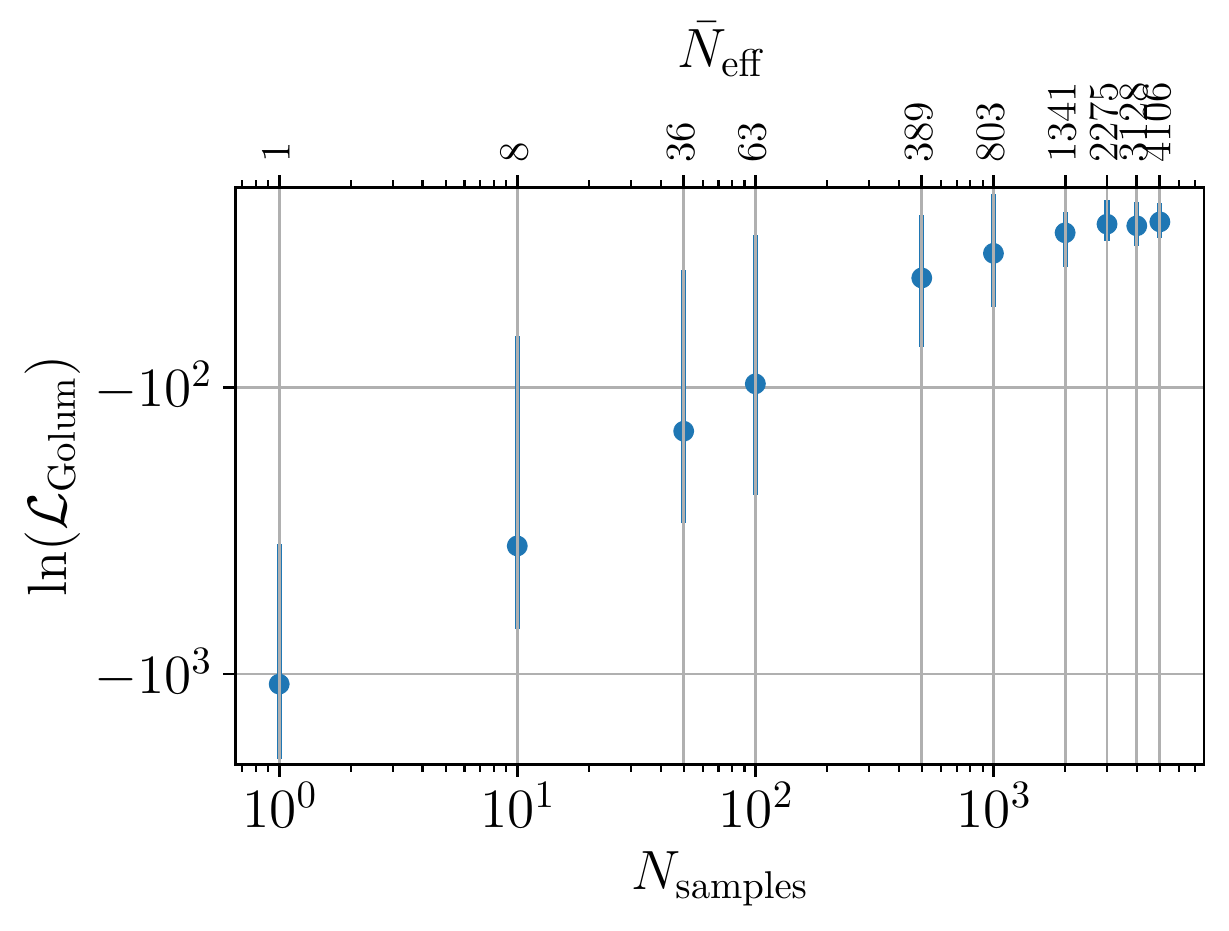}
    \caption{Evolution of the variability on the likelihood in \golum as a function of the number of samples and the corresponding mean number of effective samples. More samples lead to better stability from one iteration to the other. We reach stability once we are above 1000 effective samples.}
    \label{fig:Nsamp_Likeli_stab}
\end{figure}

Therefore, when analyzing a large number of events, it is recommended to choose the number of effective samples one wants to use and adjust the number of sub-samples taken from the first image posterior for each pair accordingly, ensuring consistency from one analysis to the other. The option to compute the number of effective samples using Eq.~\eqref{eq:n_eff_est_gol} has been adapted to the software~\citep{Golum_git}.

\section{Adapting GOLUM to low-latency}
\label{sec:low_lat_gol}
Using the lookup table presented in Sec.~\ref{sec:gol_speed}, the second image run takes $\mathcal{O}(5 \, min)$ for 1000 samples, and about $\mathcal{O}(20 \, min)$ for 5000 samples applying the same experimental conditions as in~\citet{Janquart:2021qov}. This is to compare with $\mathcal{O}(1\,hr)$ for 1000 samples in \golum's previous implementation. Therefore, the second image run is much closer to the posterior overlap computation time, and the \golum run time is dominated by the first image analysis, taking the same computational time as traditional single BBH parameter estimation. However, using the distributed evidence~\eqref{eq:distributed_joint_evidence}, the coherence ratio~\citep{Janquart_2022}
\begin{equation}\label{eq:simplify_clu}
    \clu = \frac{p(d_2 | d_1, \lenshypo)}{p(d_2 \ \unlhypo)}\frac{p(d_1 | \lenshypo)}{p(d_1 | \unlhypo)} \approx \frac{p(d_2 | d_1, \lenshypo)}{p(d_2 \ \unlhypo)} \, ,
\end{equation}
where the ratio of evidence under the lensed and the unlensed hypotheses for $d_1$ is approximated at one. This approximation is valid only when no strong HOM contributions are present in the data. This is generally the case for the current observation. Indeed, up to now, only two BBH events have been detected with a significant HOM contribution: GW190412~\citep{LIGOScientific:2020stg} and GW190814~\citep{LIGOScientific:2020zkf}, hence about one event out of 50. Moreover, it is not enough to have HOMs in the data. Their impact on the $\clu$ would be significant only if one has a type II image, leading to a significant degeneracy lifting between the Morse phase and the signal's phase~\citep{Ezquiaga:2020gdt, 2022arXiv220206334V, Janquart:2021nus}. This can potentially lead to the detection of the image type~\citep{Wang:2021kzt, 2022arXiv220206334V, Janquart:2021nus}, and biases in the recovered posteriors~\citep{2022arXiv220206334V}. However, looking attentively into the details of~\citet{2022arXiv220206334V},  the difference in evidence between lensed and unlensed events for second-generation detectors is large only for specific events, with a very low mass ratio and high total mass. Such systems are quite unlikely to be observed according to current population models~\citep{2021arXiv211103634T}.  As a consequence, the approximation above is generally valid, and one can neglect the evidence ratio for the first image.

We also show this in more detail ourselves. We injected 100 BBHs, selected from the observed mass and redshift distributions given in~\citet{2021arXiv211103634T} and with a network SNR higher than 8 for a network of two LIGO detectors~\citep{TheLIGOScientific:2014jea} and the Virgo detector~\citep{TheVirgo:2014hva} at design sensitivity. We then give an extra Morse phase corresponding to a type II image to all signals and inject them into noise. We then do the recovery under $\lenshypo$, hence accounting for a Morse factor, and once under $\unlhypo$, hence not accounting for the Morse factor. We account for HOMs and precession by injecting and recovering the signals using the \textit{IMRPhenomXPHM} waveform~\citep{Pratten:2020ceb}. Fig.~\ref{fig:ev_diff_typeII} shows the difference in evidence under the two hypotheses. The unlensed run uses \textsc{Bilby}~\citep{Ashton:2018jfp}, while the lensed runs use~\golum. In the two cases, we use the \textsc{Dynesty} sampler~\citep{Speagle:2020nld}. All the values are distributed around zero, showing that the approximation is valid in most cases when following the expected BBH population. Out of all the events, only three have a deviation of more than one in $\ln(\mathcal{Z})$, which are events with more prominent HOMs. Looking at the coherence ratios found in other studies for lensed events~\citep{Janquart_2022}, one sees lensed events typically have $\ln(\clu) \geq 5$. Therefore, a lensed event should not be discarded with this approximation, even if it has some non-negligible HOM contribution. This was already hinted at by~\citet{Janquart_2022}. We also note that outliers are possible and HOMs should formally be accounted for. Even if they would probably not modify the coherence ratio too much, one should treat them cautiously. Checking the results using the first image run would be helpful in such a case. Still, to follow the pace of the detections in low-latency, neglecting the first image can be a valuable trick.

\begin{figure}
    \centering
    \includegraphics[keepaspectratio, width=0.49\textwidth]{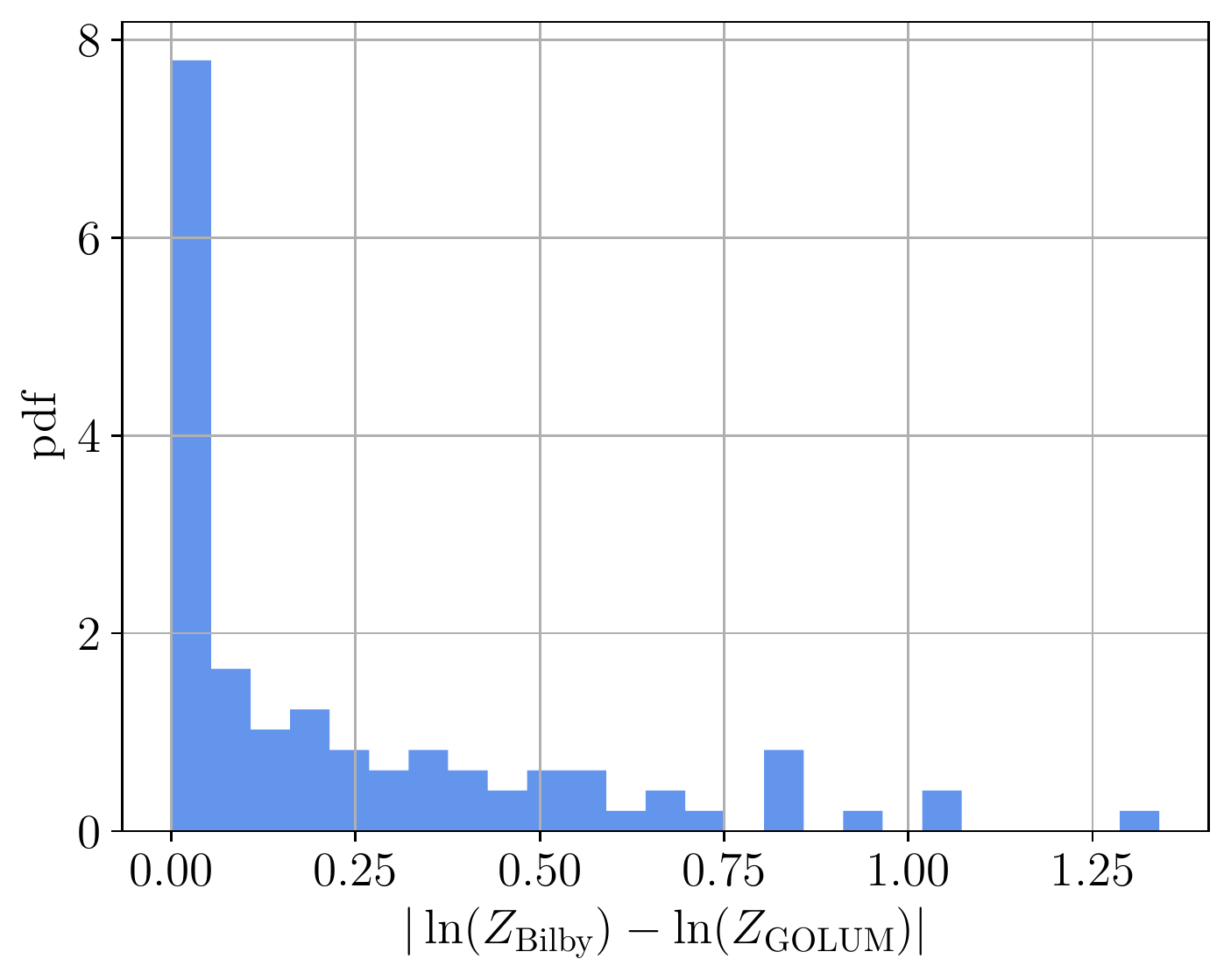}
    \caption{Difference between the log evidence recovered for type II images using \textsc{Bilby} and \golum. One sees that most of the evidence is well below 1, meaning that the error made is not big enough to miss a lensed pair due to this approximation.}
    \label{fig:ev_diff_typeII}
\end{figure}

Posterior overlap~\citep{Haris:2018vmn} also assumes that the posteriors obtained through the unlensed analyses are unbiased and representative of the ones obtained under the lensed hypothesis. An event breaking the approximation made in Eq.~\eqref{eq:simplify_clu} would also break the approximation made in the posterior overlap analysis. Therefore, our suggested method is not necessarily more keen on issues when confronted with HOMs. On the other hand, our framework still accounts for more correlation between the events since we do not have to perform KDE reconstruction or take a sub-part of the parameter space. Therefore, using this fast \golum should reduce the number of false alarms and, thus, the number of events that need to be followed up by more complete joint parameter estimation methods.

We also note it should be possible to convert the unlensed results into lensed ones using hybrid sampling~\citep{2022arXiv220812872W}, making it possible to change the dimensionality of the problem and rapidly correct for the strong-lensing effects. This could make it possible to use the exact \golum framework in low-latency. This is left for future work.

\section{Symmetric GOLUM}
\label{sec:symm_gol}

One remaining possible issue is the parameter space coverage. In principle, one expects the posteriors for the two events to match perfectly. However, in reality, noise effects, the number of online detectors, etc. can lead to variations in the posteriors with changes in the widths and shifts between the posteriors. Because of that, \golum can sometimes give slightly different posteriors than a joint parameter estimation tool. This does not mean we miss the correct value, but more that we can get overconfident in the final posteriors and miss a part of the lower significance region. An illustration of this phenomenon is given in Fig.~\ref{fig:IllustrationSampling}, where we illustrate the different posteriors as simple Gaussians. When considering only one of the two images, the final samples can miss a small part of the region obtained for the joint analysis.  Once we have acquired the posteriors for the first image, \golum uses them to probe the lensing parameters, but they are not re-sampled. Therefore, we cannot create new samples, and regions not covered by the samples taken for the first image are not considered in the joint distribution. However, it can be the case that the joint parameter estimation extends to a region compatible with the second image but not covered by the first image\footnote{The effect can also be observed in Fig.9 of~\citet{Lo:2021nae}.}.

\begin{figure}
    \centering
    \includegraphics[keepaspectratio, width=0.49\textwidth]{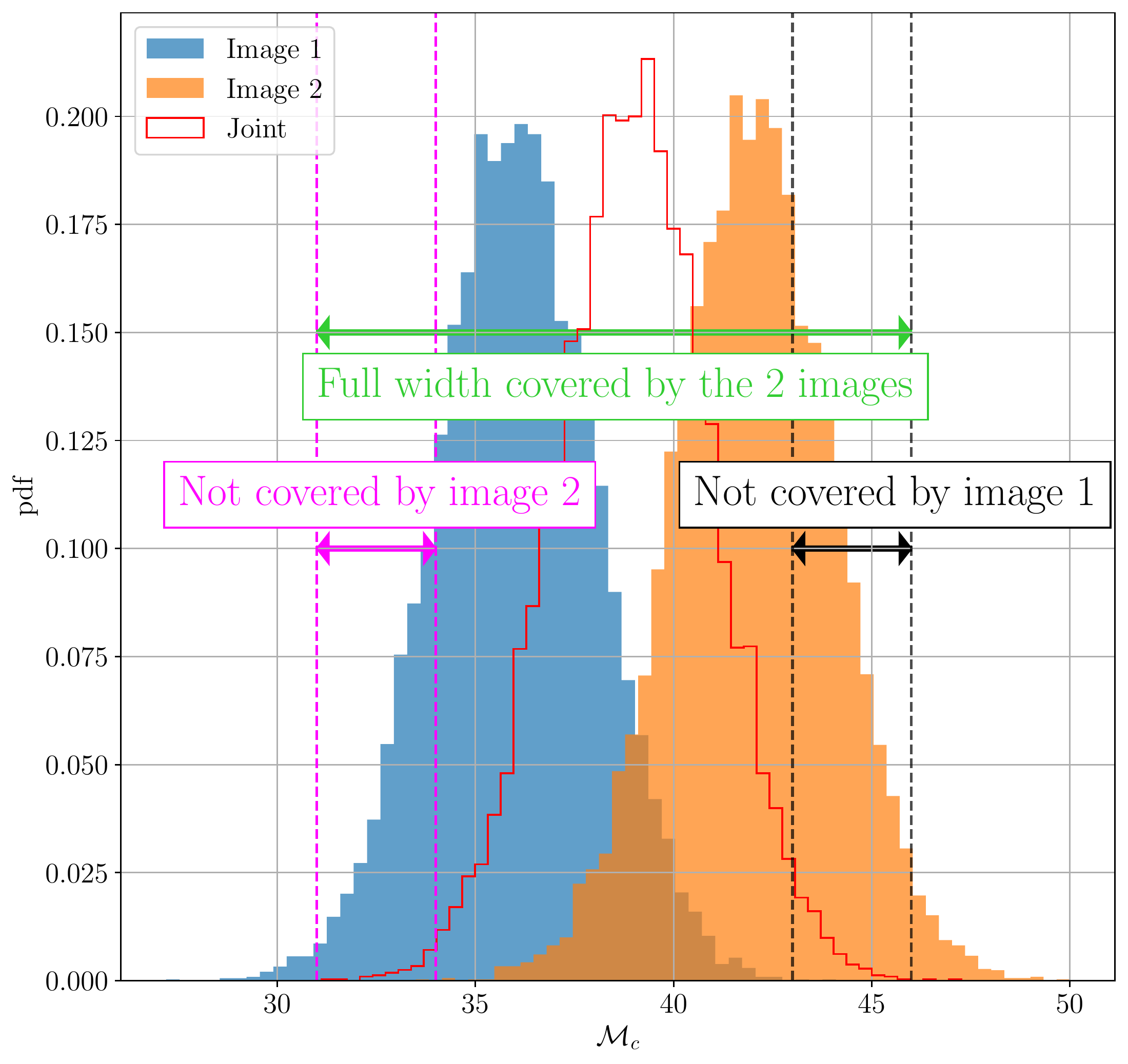}
    \caption{Illustration of the coverage of the joint parameter space for fiducial chirp mass recoveries. We straightforwardly use Gaussians to show the effect. Sampling only one of the two images is not enough to fully cover the joint results. However, using the two images works, motivating the use of a symmetric version of \textsc{GOLUM}.}
    \label{fig:IllustrationSampling}
\end{figure}

So, while in theory, the labels 1 and 2 in Eq.~\eqref{eq:distributed_joint_evidence} should not matter, they do have a consequence on the final result due to the difference in observation condition. However, one can easily show that 
\begin{align}\label{eq:symm_evidence}
    p(d_1, d_2 | \lenshypo) &= p(d_2 | d_1, \lenshypo)p(d_1 | \lenshypo) \nonumber \\ &= p(d_1 | d_2, \lenshypo) p(d_2 | \lenshypo) \nonumber \\
    &= \frac{1}{2}\bigg( p(d_2 | d_1, \lenshypo)p(d_1 | \lenshypo) \nonumber \\
    & \quad + p(d_1 | d_2, \lenshypo) p(d_2 | \lenshypo) \bigg) \, ,
\end{align}
where the conditional likelihood is re-expressed simply as the conditional of $d_1$ given $d_2$. Eq.~\eqref{eq:symm_evidence} is symmetric in the two events. Therefore, the parameter space is explored by the two events. For any of the two terms, the conditional evidence is computed using Eq.~\eqref{eq:cond_ev_marg_likeli}, once using the samples from image 1 to analyze image 2, and once using samples from image 2 to analyze image 1\footnote{Image 1 and 2 here are just arbitrary labels to make the difference between them.}. So, this conditioned likelihood now requires two usual BBH runs (accounting for the Morse factor) and two \golum runs. Since the runs can be done in parallel, the total time of the run is the time needed to analyze a single BBH. In this case, the coherence ratio takes the more complex form
\begin{align}\label{eq:symme_clu}
    \clu &= \frac{1}{2}\frac{p(d_2 | d_1, \lenshypo)p(d_1 | \lenshypo) + p(d_1 | d_2, \lenshypo) p(d_2 | \lenshypo)}{p(d_1 | \unlhypo)p(d_2 | \unlhypo)} \nonumber \\
    &= \frac{1}{2}\Bigg[ \frac{p(d_2 | d_1, \lenshypo)}{p(d_2 | \unlhypo)}\frac{p(d_1 | \lenshypo)}{p(d_1 | \unlhypo)} + \frac{p(d_1 | d_2, \lenshypo)}{p(d_1 | \unlhypo)}\frac{p(d_2 | \lenshypo)}{p(d_2 | \unlhypo)} \Bigg] \nonumber \\
    &\simeq \frac{1}{2} \Bigg[ \frac{p(d_2 | d_1, \lenshypo)}{p(d_2 | \unlhypo)} + \frac{p(d_1 | d_2, \lenshypo)}{p(d_1 | \unlhypo)} \Bigg] \, ,
\end{align}
where the last line assumes a low HOM hypothesis, as detailed in Sec.~\ref{sec:low_lat_gol}. In this case, only two \golum runs are needed, leading to something close to joint parameter estimation in less than an hour. 

In addition to having evidence better accounting for the parameter space covered by the two images, one is generally also interested in the posteriors from the joint analysis. The posteriors can be expressed as
\begin{align}\label{eq:symm_posts}
    p(\bvarTheta, \boldPhi | d_1, d_2) &= p(\bvarTheta | d_1, d_2)p(\boldPhi | d_1, d_2) \nonumber \\
    &= \frac{1}{2} \bigg( p(\bvarTheta | d_1, d_2)p(\boldPhi | d_1, d_2) \nonumber \\
    &\qquad + p(\bvarTheta | d_1, d_2)p(\boldPhi | d_1, d_2) \bigg) \, .
\end{align}
Each $p(\bvarTheta | d_1, d_2)$ in Eq.~\eqref{eq:symm_posts} can be expressed using Eq.~\eqref{eq:simple_reweighitng}. In our context, one of the terms is computed using $p(\bvarTheta\ | d_1)$ and $p(d_2 | \bvarTheta, \boldPhi)$ and the other is obtained by swapping images 1 and 2 in the computation. Hence, we account for $\bvarTheta$ samples coming from the two posteriors, and we more largely cover the parameter space. This avoids us from missing a part of the parameter space because of narrower priors in the first image analysis. 

This process pushes \golum closer to formal joint parameter estimation while not significantly increasing the computational time, especially when accounting for the improvements detailed in Sec.~\ref{sec:gol_speed}.

\section{Confrontation with joint parameter estimation}
\label{sec:compa_jpe}

In this section, we compare the results from the \golum analyses using the above-mentioned updates. We require 2000 effective samples. The 15 analyzed BBHs are generated following the rates and population models given in~\citet{2021arXiv211103634T}. The lensing parameters linking the two events are sampled from the expected distribution for an SIE lens model~\citep{More:2021kpb}. We use a uniform prior between 5$M_{\odot}$ and 100$M_{\odot}$ for the chirp mass ($\mathcal{M}_c = (m_1 m_2)^{3/5}/(m_1 + m_2)^{1/5}$) and between 0.1 and 1 for the mass ratio ($q = m_2/m_1$). The individual masses are constrained between 1 and 1000$M_{\odot}$. For the other parameters, we use the usual priors. When considering individual Morse factors, we use a discrete prior on the three possible values, while for the Morse factor difference, we consider $n_{21} \in \{0, 0.5, 1, 1.5\}$. The prior on the relative magnification is uniform between 0.1 and 50, while the prior on the time delay is $\mathcal{U}(t_{21}^{\mathrm{inj}} - 0.2, t_{21}^{\mathrm{inj}} + 0.2)$. The events are injected in a network made of the two LIGO detectors and the Virgo detector at design sensitivity. We require the network SNR to be higher than 8 for the two images\footnote{Application of \golum to sub-threshold pairs was illustrated in~\citet{Janquart:2021qov}.} and perform the analysis. The unlensed runs are done using \textsc{Bilby}~\citep{Ashton:2018jfp} with the \textsc{Dynesty} sampler~\citep{Speagle:2020nld}. The joint likelihood analysis is performed using the framework added into \golum~\citep{Golum_git}, and the distributed runs use the updated version of \golum explained above, using the same sampler. Again, the injections and analyses are performed using the \textit{IMRPhenomXPHM} waveform~\citep{Pratten:2020ceb}.

First, in Fig.~\ref{fig:likeli_compa_jpe}, we compare the difference in likelihood for the symmetric \golum framework and the joint parameter framework and the old \golum implementation. With the new design, we obtain a difference in evidence between joint parameter estimation and \golum below the typical error on the evidence made by \golum\footnote{Since we sum contributions from different runs, there is an error cumulation in \golum leading larger uncertainties on the final value.}.  So, there is a real improvement in the stability of the framework obtained through symmetrization. Looking at the results for the old implementation, one sees it also performs relatively well, and can be an effective filter. We also note that, in the first approximation, one should consider as a lensed candidate all events with a coherence ratio superior to a threshold value (typically, 2 is a conservative value~\citep{Janquart_2022}). Even with the largest difference, the \golum approximate version using only one image is not discarding any lensed event. So, it can be used as a filtering algorithm and is indicative when studying a large-scale population. 

\begin{figure}
    \centering
    \includegraphics[keepaspectratio, width=0.49\textwidth]{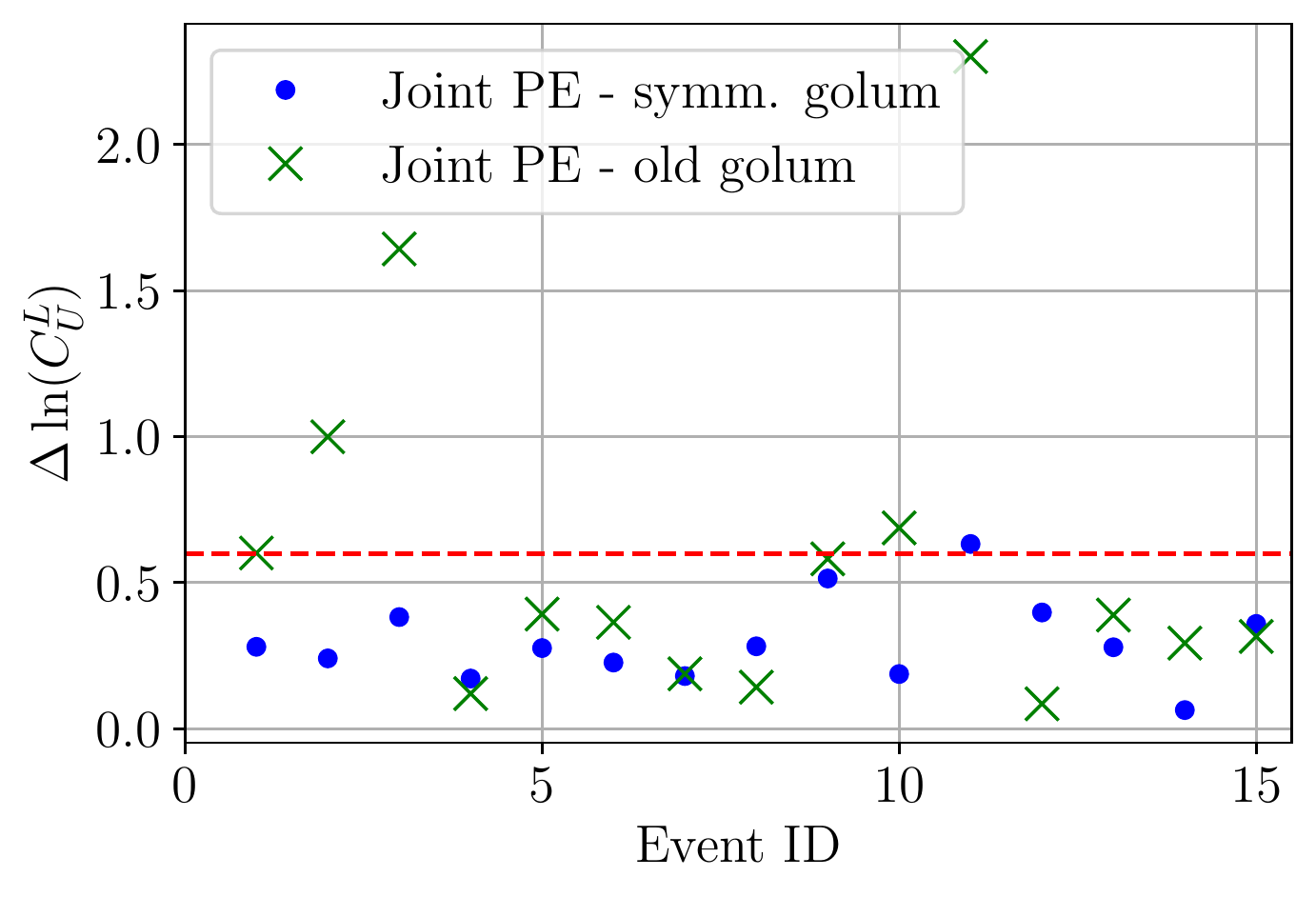}
    \caption{Comparison between the error on the evidence obtained through the symmetric \golum implementation and the old implementation. The 1-image implementation already performs sensibly well. However, there is an improvement when considering the symmetric version, where nearly all the differences are below the evidence error present in \golum. }
    \label{fig:likeli_compa_jpe}
\end{figure}

A second problem tackled by the use of symmetrization is posterior coverage. The main idea is that it enables one to cover the parameter space better than when using only one image. One of the issues found with the usual \golum is that it sometimes does not fully cover the parameter space. Jointly, the sampling process has inherent issues, such as vanishing samples. The symmetric implementation naturally addresses these issues. As represented in Fig.~\ref{fig:IllustrationSampling}, it helps in covering the entire parameter space. Secondly, since it uses two \golum runs, more independent samples are directly available, and one reduces the issue of vanishing samples. Therefore, we get a better match between the symmetric \golum posteriors and those coming from joint parameter estimation. This is shown in Fig.~\ref{fig:sky_loc_rec}, top panel, where we represent the sky location recovery. Another issue that can arise, and only partially solved with symmetrization, is the domination of a reduced number of samples in the reweighting process. This can happen when a couple of data points have weights $p(d_i | \bvarTheta, \boldPhi)/p(d_i, d_j | \boldPhi)$ significantly higher than the others. In such a case, the selection process latches on them, and one gets scattered posteriors with biased dominant regions, leading to high-density points. This is less likely to happen when considering more samples, which is why symmetrization is helpful. However,in some cases, scattering is still possible. The dominant value may be slightly away from the actual injected ones for some parameters. Therefore, the injected value is not in the highest density region. The posteriors are then more different between joint parameter estimation and \golum. Still, we note that, with symmetrization, we have not seen cases where the injected value had no samples when the joint parameter estimation had. Several possibilities to correct this exist. First, one could simply use more samples. However, covering the full space requires testing millions of combinations, which becomes computationally expensive. On the other hand, since there are samples in the correct region, one can think about using importance sampling using the formal joint parameter estimation likelihood. An example of a case where the biased higher density region leads to different posteriors between joint parameter estimation and \golum is given in Fig.~\ref{fig:sky_loc_rec}, bottom panel. While the injected value is not in the 90\% confidence interval, it is in the 95\% one. The scattered posteriors problem was also present in the old \golum framework and was more pro-eminent. So, even if it does not completely solve the issue, symmetrization helps in significantly reducing it. 

\begin{figure}
    \centering
    \includegraphics[keepaspectratio, width=0.45\textwidth]{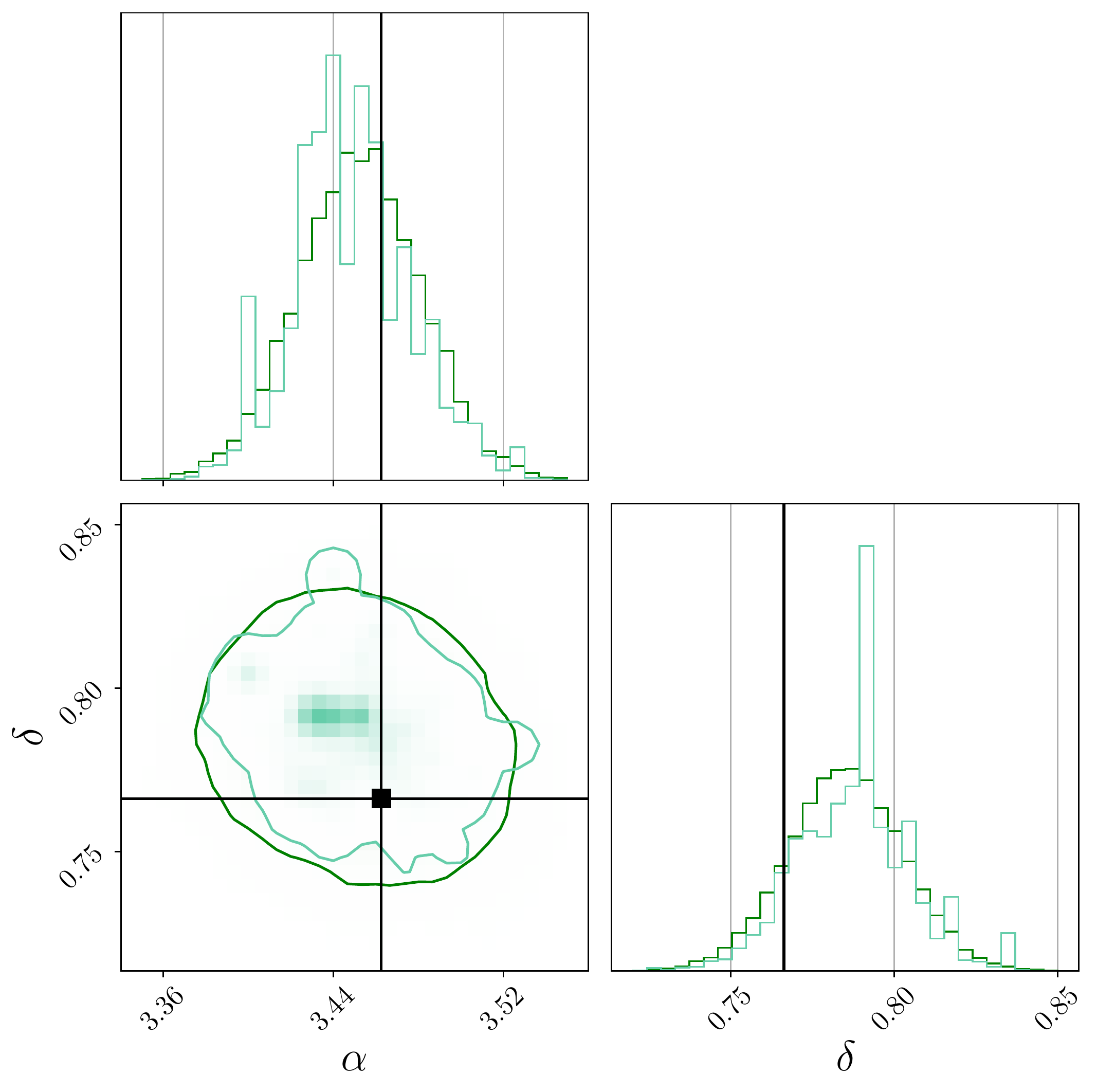}
    \includegraphics[keepaspectratio, width=0.45\textwidth]{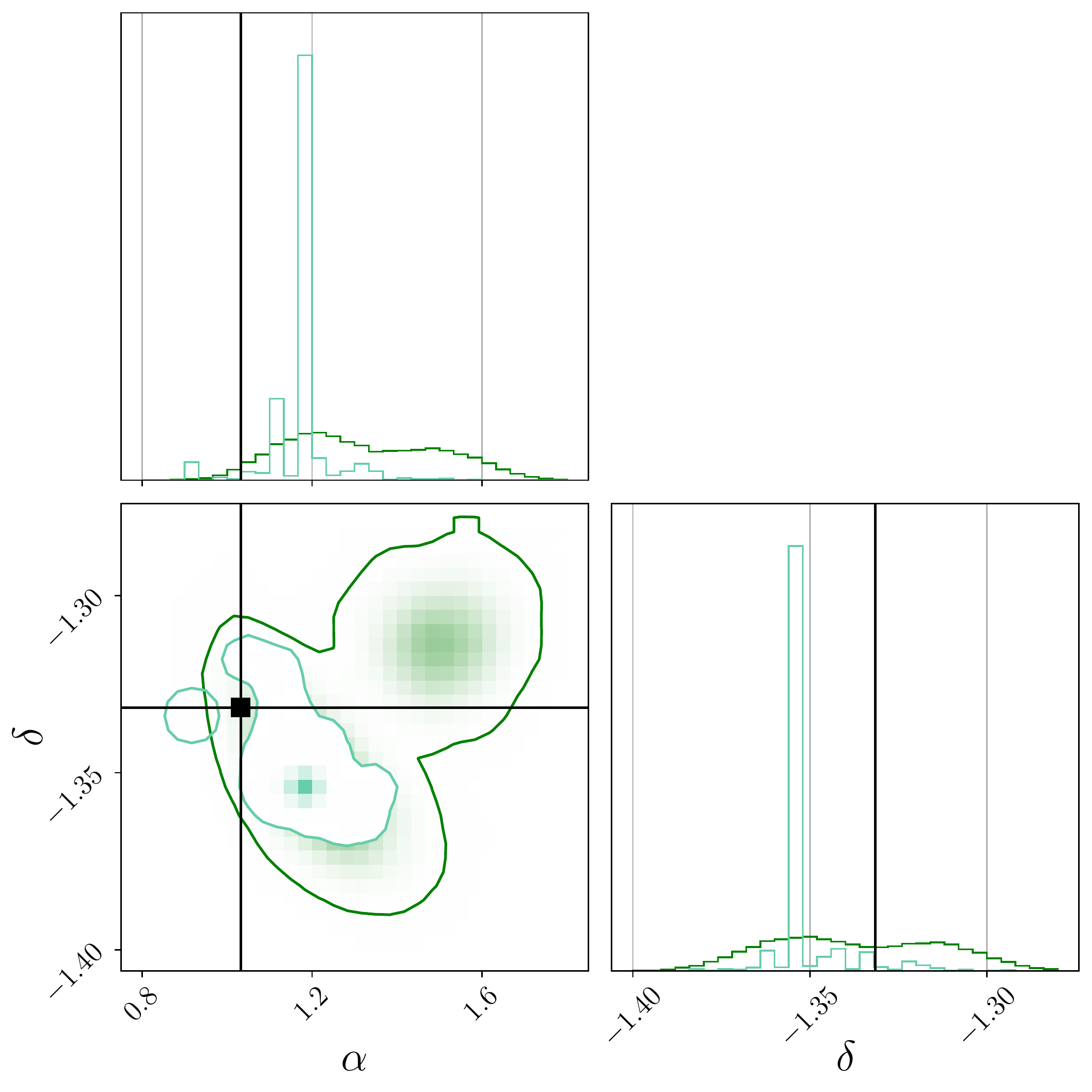}
    \caption{Sky recovery for the symmetric \golum framework (light blue) and complete joint inference (dark green). \emph{Top:} \golum and joint parameter estimation have matching 90\% confidence interval recovery. \emph{Bottom:} Scenario in which there are dominating samples in the reweighting process leading to a high-density region peaking away of the 90\% interval, artificially shrinking the region and exclusing the injected value. One can also see the clear presence of dominant samples in the 1D posteriors, where there are a couple of bins with much more importance. This issue is reduced when using symmetric \golum.}
    \label{fig:sky_loc_rec}
\end{figure}

\section{Confrontation with real data}
\label{sec:o3_data}

The previous sections have shown symmetrization helps in getting \golum closer to the actual joint parameter estimation, and that one can generally skip the first image run to save time. In this section, we use the symmetric \golum and the publicly available data~\citep{ligo_scientific_collaboration_and_virgo_2021_5546663} from the GWTC-3 LVK catalog~\citep{LIGOScientific:2021djp}. We use the public samples and evidence obtained with \textsc{Bilby}~\citep{Ashton:2018jfp} and the \textsc{IMRPhenomXPHM}~\citep{Pratten:2020ceb} waveform as results for the unlensed analysis. We also assume the HOMs to be negligible for all the events and do the approximate symmetric \golum analysis. To make a fair comparison, we also reweight all the posteriors and evidence from the initial data to correspond to values with the same priors. This avoids us from favoring some events because of the initial prior volume. In total, 87 BBHs are considered hence we analyze 3741 event pairs. The histogram of the $\ln(\clu)$ values, zoomed on the -100 to 14 value region, is presented in Fig.~\ref{fig:lnClus_O3}. There is a tail of events going down to a few hundred, but it is not so informative. We see that the vast majority of the events are well below zero. Only 43 event pairs have a positive coherence ratio. Out of those, most have $\ln(\clu) < 5$. 

\begin{figure}
    \centering
    \includegraphics[keepaspectratio, width=0.49\textwidth]{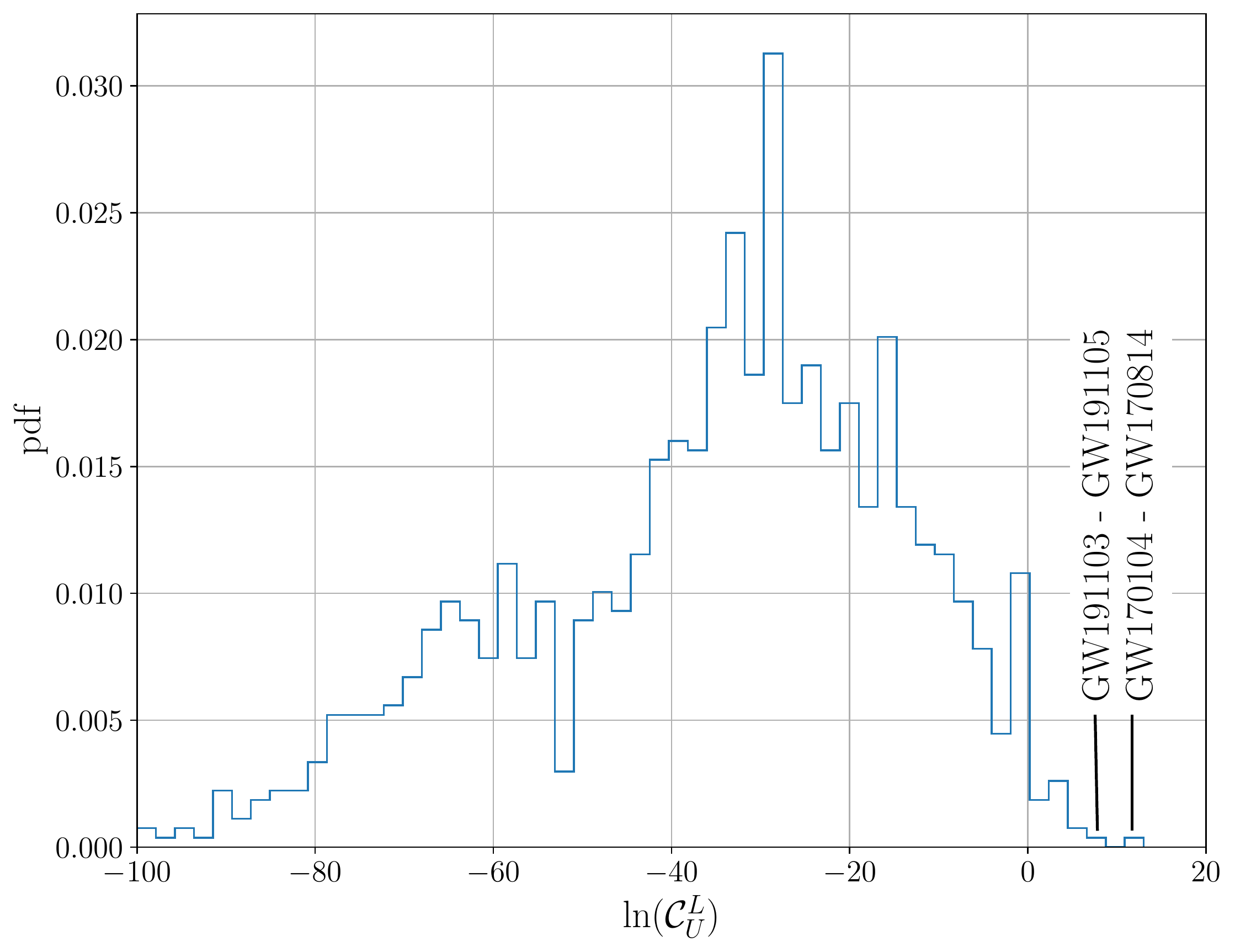}
    \caption{Histogram of the $\ln(\clu)$ values found for all the GW events using the symmetric \golum approach. Most of the events have a coherence ratio below zero. We also recover the different intriguing candidates reported in the literature: GW191103-GW191105 and GW170104-GW170814. While probably not lensed, recovering them shows that our method is trustworthy.}
    \label{fig:lnClus_O3}
\end{figure}

We note, however, the presence of two particular events in the higher tail, matching previously reported events. First, the GW191103-GW191105 event pair has a relatively high coherence ratio, $\ln(\clu) = 6.012$. This event is also reported in~\citet{LIGOScientific:2023bwz}, where it is seen as an interesting candidate due to its characteristics. It also has a time delay of only about two days and a relative magnification close to 1. In addition, the two images are likely to be the same type. Having two type-I images close to each other is a probable scenario for galaxy-lensing~\citep{More:2021kpb}. Therefore, the values are compatible with the lensing models. However, \citet{LIGOScientific:2023bwz} shows that it is not lensed when accounting for the population models. The posterior distribution we obtain for the relative magnification and the difference in Morse factor are represented in Fig.~\ref{fig:lensing_params_11031105}.

\begin{figure}
    \centering
    \includegraphics[keepaspectratio, width=0.49\textwidth]{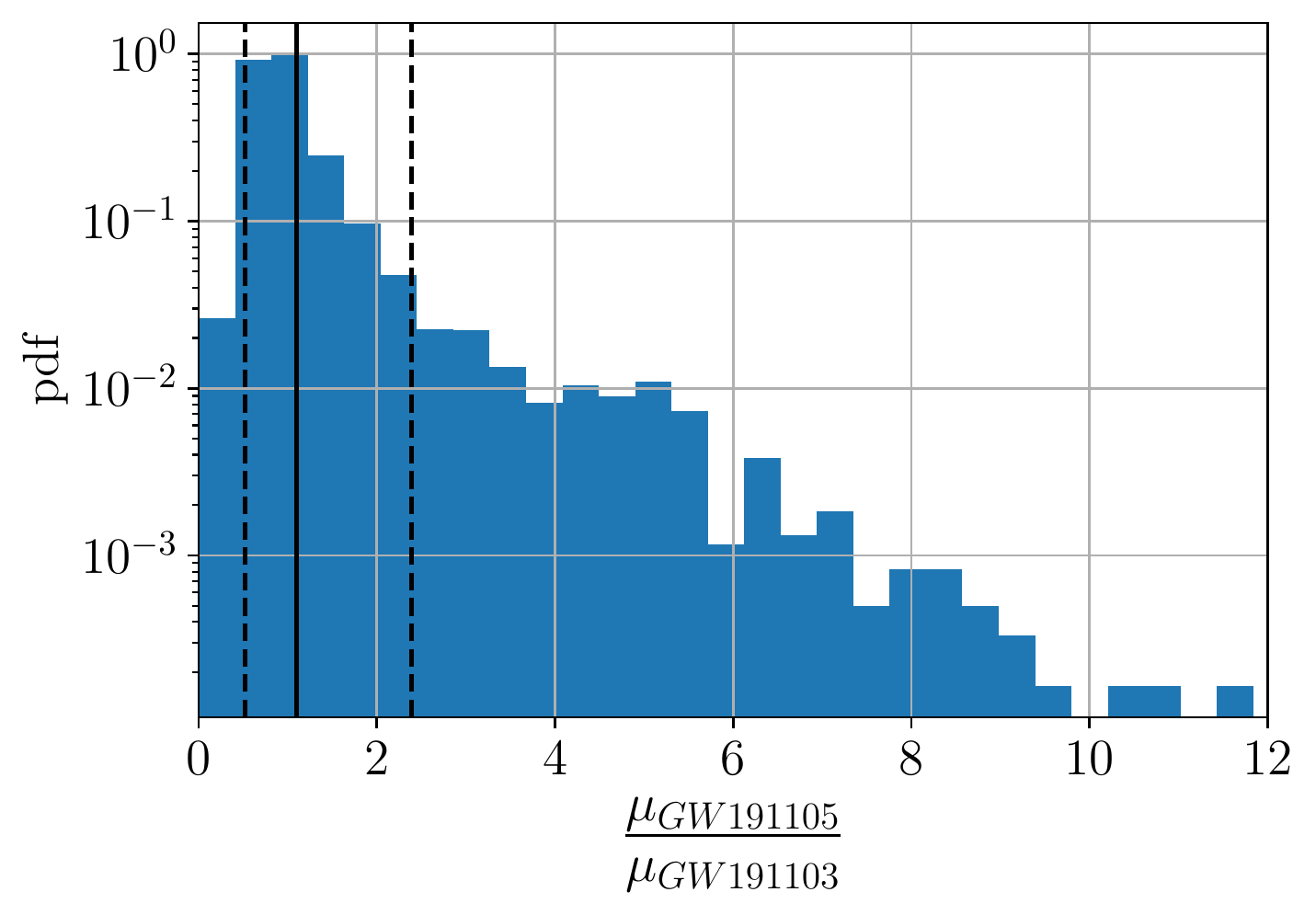}
    \includegraphics[keepaspectratio, width=0.49\textwidth]{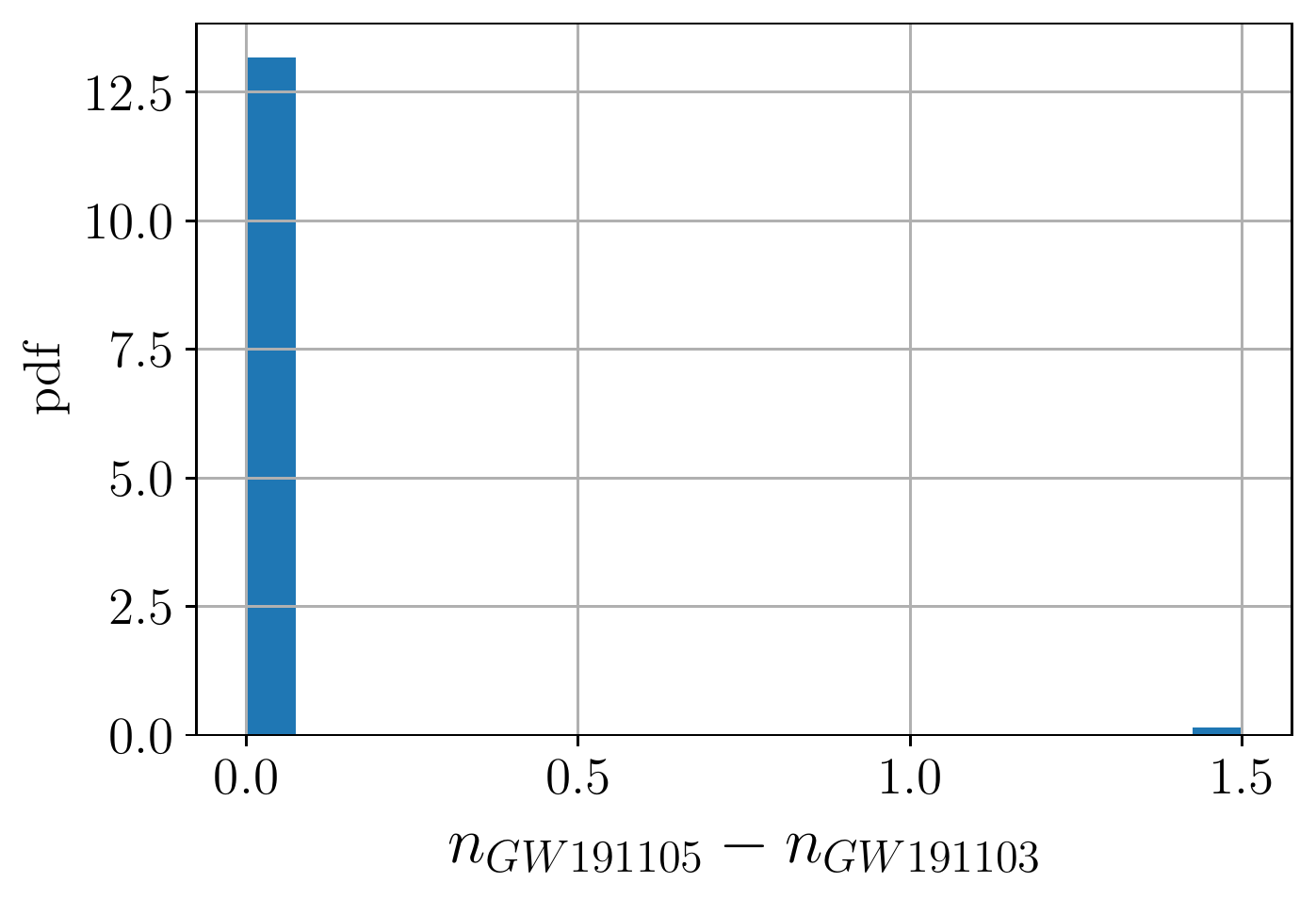}
    \caption{\emph{Top:} Posterior recovery for the relative magnification for GW191103-GW191105. \emph{Bottom:} Posterior recovery for the difference in the Morse factor. The two events also have a time delay of $\sim 2.5$ days. Therefore, the recovered parameters are compatible with the expectation for a galaxy lens~\citep{More:2021kpb}.}
    \label{fig:lensing_params_11031105}
\end{figure}

The other high value found is for GW170104 - GW170814, where $\ln(\clu) = 11.153$, which is much higher than all the others. This event was already reported as significant in the literature~\citep{Hannuksela:2019kle, Dai:2020tpj,Liu:2020par}. However, it has a long time delay (we find 222.01 days) and a large  difference in magnifications ($\mu_{\mathrm{GW170814}-\mathrm{GW170104}} = 0.37^{+0.08}_{-0.07}$, such that $D_L^{\mathrm{GW170814}} = \sqrt{\mu_{\mathrm{GW170814}-\mathrm{GW170104}}} D_L^{\mathrm{GW170104}}$). In addition, we find that the Morse factor difference between the two events is 1. Therefore, one of the two images is a type III image. This is not compatible with the expectations for galaxy lenses (see~\citet{Collett:2015tma, Collett:2017ksf, Wierda:2021upe, More:2021kpb}, for example). This situation could be a bit more likely for a galaxy-cluster lens, but this type of lensing has a low probability (see, for example,~\citet{Smith:2017mqu, Smith:2018gle, Smith:2019dis, Robertson:2020mfh, Ryczanowski:2020mlt}). Therefore, it seems unlikely for the event to be lensed and we do not push its study further in this work.

We do not report any new detections or claims. However, we have shown that our method can reproduce the results presented in literature over the years on a much shorter time scale, owing to its enhanced speed. 

\section{Conclusions}
\label{sec:conclusions}

In this work, we presented upgrades to the \golum framework, a tool to analyze strongly-lensed gravitational waves. We have shown how recasting the lookup table could reduce memory usage and computation time and how computing the number of effective samples leads to better stability. We then showed how the pipeline can be used in low latency by showing the generally small impact of the first image contribution to the coherence ratio. Finally, we have introduced a symmetrization of the pipeline, solving some issues found in the framework, such as difficulties in fully covering the parameter space when using only one image and vanishing samples in the reweighing process. 

For each of the upgrades, we have shown how they work and what their effects are. In particular, we have compared the performances of the entire symmetric \golum framework with results coming from joint parameter estimation, also implemented in~\citet{Golum_git}. We started by showing that, generally, the first image run does not impact too much the evidence computation. Therefore, one is not forced to do it and can use the samples from the unlensed searches. Since the most computationally-heavy part of \golum is the first image evaluation, dropping the first image analysis significantly reduces the computation time. In addition, the run time for the second image has been improved by a factor of ten thanks to the different upgrades, making \golum quite suited for low-latency applications. This approach contains a certain degree of risk, as it can fail if important HOM contributions are present in the data. However, this would be flagged in the usual analyses, making the approximation evident. We also note that other low-latency methods, like posterior overlap, do not account for the Morse factor and are, consequently, sensitive to HOMs.

Using symmetrization leads to a clear improvement in the evidence compared to a single image, with values much closer to those given by joint analyses. In addition, we have shown that we get a better sample coverage with less scattered sky recovery. A remaining issue, only partially solved via the symmetric run, is the domination of a few samples in the weights. Indeed, a reduced number of samples can have much higher probability ratios in the reweighting process, leading to an artificial high-density region in the final posterior. This problem is already significantly reduced when using symmetrization, but there are still some cases where the final posterior is biased due to dominant weights. We suggest two possible avenues to solve this: use more samples to cover the parameters space better or perform importance sampling to correct for the bias using the formal joint likelihood. The main issue is that such methods are computationally heavy when applied to numerous samples. Still, it already gives a good idea about the sky localization of the event in case one would like to perform a follow-up analysis. 

Finally, we have re-analyzed all the GW BBH candidates released by the LVK since the first detection. Using the public samples and set-up, we performed (approximate) symmetric \golum runs for all the possible pairs. Our results match those of the LVK collaboration and other results given in the literature. Most events are certainly unlensed, and there are two notable event pairs: GW191103-GW191105 and GW170104-GW170814. The first has the second highest $\clu$ but also has the particular property of having apparent lensing parameters compatible with the current galaxy lens models. The O2 pair is the one with the highest coherence ratio in all the pairs: $\ln(\clu) = 11.153$. While intriguingly high, other works have shown why this event pair is probably not lensed. 

In the end, we have shown how improving our pipeline enables it to be faster and more reliable, making it possible to analyze large chunks of data in an approximate (but still accurate) form. This is a crucial step for future studies, where the number of events will grow steadily, and more and more pairs will have to be tracked. Therefore, using a more precise method with comparable speed is an important step forward to reduce the false-alarm probability. Let us also note that, with this approach, only a few events would have to be followed up using joint parameter estimation and low-latency results could be issued quickly in case an exceptional event is detected. 

\section*{Acknowledgements}
\label{sec:Acknow}
The authors thank Harsh Narola, Sylvia Biscoveanu, and John Vietch for useful discussions. We also thank Mick Wright for carefully re-reading the manuscript.
J.J, K.H, and C.V.D.B. are supported by the research program of the Netherlands Organisation for Scientific Research (NWO).
The authors are grateful for computational resources provided by the LIGO Laboratory and supported by the National Science Foundation Grants No. PHY-0757058 and No. PHY-0823459. This material is based upon work supported by NSF's LIGO Laboratory which is a major facility fully funded by the National Science Foundation.  

\section*{Data availability}
The data underlying this article will be shared in reasonable request to the corresponding authors.

\bibliography{bibly}

\end{document}